\documentclass[prl,twocolumn,amsmath,amssymb,superscriptaddress,nofootinbib]{revtex4-2}
\usepackage{graphicx}
\usepackage{amsmath}
\usepackage{bm}
\usepackage{xcolor}
\usepackage{epstopdf}
\usepackage{subfigure}
\usepackage{hyperref}

\begin{document}

\title{Zero-energy Andreev bound states in iron-based superconductor Fe(Te,Se)}

\author{Zhe Hou}
\affiliation{Department of Physics, University of Basel, Klingelbergstrasse 82, CH-4056 Basel, Switzerland}
	
\author{Jelena Klinovaja}
\affiliation{Department of Physics, University of Basel, Klingelbergstrasse 82, 
		CH-4056 Basel, Switzerland}

\begin{abstract}
Majorana bound states have been predicted to exist in vortices of topological superconductors (SC). A realization of the Fu-Kane model, based on a three-dimensional topological insulator brought into proximity to an $s$-wave SC, in iron-based SC Fe(Te,Se) has attracted strong interest after pronounced zero-energy bias peaks were observed in several experiments.
Here, we show that, by taking into account inhomogeneities of the chemical potential or the presence of potential impurities on the surface of Fe(Te,Se), the emergence of these zero-energy bias peaks can be explained by trivial Andreev bound states (ABSs) whose energies are close to zero. Our numerical simulations reveal that the ABSs behave similarly to Majorana bound states. ABSs are localized only on the, say, top surface  and cannot be distinguished  from their topological counterparts in transport experiments performed with STM tips. Thus, such ABSs deserve a careful investigation of their own.
\end{abstract}

\maketitle

\emph{Introduction.} 
Majorana bound states (MBSs) attracted substantial interest due to their promise to open a path to realize topologically protected operations for fault-tolerant quantum computing based on their non-Abelian braiding statistic~\cite{ReadGreen, Beenakker1, Flensberg, Wilczek, SCZhang1, Katharina, Ivanov, VolovikBook, BeenakkerAnnual,Sato, Biswas, MengCheng}. 
Since the first model of MBSs in one-dimensional (1D) $p$-wave superconductor (SC) was proposed by Kitaev \cite{Kitaev, Lieb} many  scenarios for generating MBSs have been put forward, e.g. in semiconductor nanowires with strong spin-orbit coupling and Zeeman field \cite{Sarma1, Oppen, Fisher, Tewari1, Loss1, Mourik, Das, Rokhinson, Deng}, in carbon-based materials \cite{Klinovaja108, Egger, KlinovajaPRX3, SauPRB88, DutreixEPJB87, MarganskaPRB97, DesjardinsNM18}, in magnetic atomic chains \cite{Yazdani1, Loss2, Simon, VazifehPRL111, ChoyPRB84, PientkaPRB88, Jack, PergeScience346, RubyPRL115, PawlakNPJ2}, 
in  topological insulator (TI) nanowires \cite{CookPRB84, CookPRB86, Legg}, 
in two-dimensional (2D) ferromagnetic insulator-semiconductor heterostructures \cite{Sarma2, Sarma3, SCZhang2, Bena, PatrickLee, SatoPRL103, SatoPRB82, SatoJPSJ85}, and in three-dimensional (3D) strong TIs coupled to conventional $s$-wave SCs \cite{Kane1, Kane2, Beenakker2, JacobPRL104, YukioPRL103, KTLawPRL103}. 
Some of the above models have been reported to be realized experimentally \cite{Mourik, Das, Rokhinson, Deng, DesjardinsNM18, PergeScience346, RubyPRL115, PawlakNPJ2, Marcus, Furdyna, JinFengJia1, JinFengJia2, Gordon, Ando, Yazdani2, Yazdani3, XuNatPhy10, XuPRL112, YinNatPhy11, ChengNatPhy16, BertholdScience364, GerboldNatCommun8, HowonSciAdv4, SujitPNAS117}, enriching the ecosphere of Majorana physics. 

Among the enumerated proposals, one of the most attractive options to be realized is the Fu-Kane model \cite{Kane1}, which utilizes the helical property of the Dirac surface states in 3D TIs as an ideal platform for producing MBSs without the  need of a Zeeman field or of fine-tuning the chemical potential inside the bulk gap. This model has been announced to be fully realized in a preexisting iron-based SC $\rm{FeTe}_{0.55}\rm{Se}_{0.45}$ by several experimental groups \cite{HongJunGao1, Shin, HongDing1, Tamegai, HongJunGao2}. It was demonstrated that the key ingredients of Fe(Te,Se) for the advent of MBSs are the intrinsic superconducting phase, originally brought out from FeSe layered system, and the strong TI phase, induced by the band inversion along the $\Gamma-Z$-line in the Brillouin zone as a result of Te substitution \cite{ZhongFang}. Such an intrinsic topological superconductivity makes Fe(Te,Se) a highly promising material for Majorana-based topological quantum computing. 

However, in spite of this progress, there are important issues to be clarified concerning this popular iron-based SC \cite{HongDing2, HaiHuWen}. Some groups have reported the low-ratio of the appearance of MBSs in the superconducting vortices of Fe(Te,Se) \cite{Tamegai}, and some reported their absence \cite{HaiHuWen}. Even though a well shaped zero-bias peak (ZBP) as well as a half-integer shifted linear scaling relation of the surface modes can be observed \cite{HongDing1}, the coexisting non-topological regions mixed with zero-energy states in the vortices on the surface might also hamper the practicability of Fe(Te,Se). Actually, this complexity in experiments can be attributed to the intrinsic chemical inhomogeneity which comes from the nonuniform substitution of Se atoms, unavoidable defects, or excess iron atoms due to the special stoichiometry of Fe(Te,Se), as validated by  scanning topography \cite{Plummer}. Regarding this ``dirty'' composition of Fe(Te,Se) SC, it is thus an urgent question to clarify whether the observed zero-energy states in Fe(Te,Se) are MBSs, which can be used as robust topological qubits, or whether this ZBP could be attributed to trivial Andreev bound states (ABSs) as was demonstrated in other MBS platforms \cite{PradaNRP2, KellsPRB86, FleckensteinPRB97, PenarandaPRB98, PtokPRB96, MoorePRB97, LiuPRB96, AlspaughPRR2, ReegPRB98, WoodsPRB100, LiuPRB99, ChenPRL123, LeePRL109, JungerPRL125, DmytrukPRB102, Yu2004, KayyalhaScience367, Valentini2008, Kim2105}.

\begin{figure}
\includegraphics[width=8.0cm, clip=]{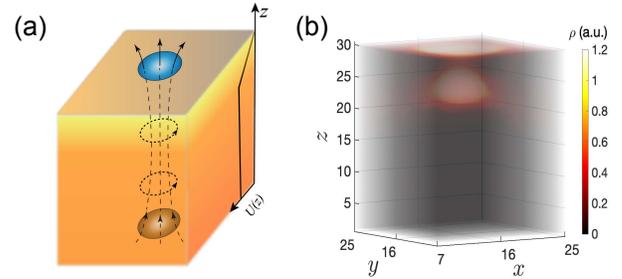}
\caption{(a) The schematic diagram of 3D TI SC  with a $z$-dependent electrostatic potential $U(z)$ shown by the solid curve. An external magnetic field induces a superconducting vortex (dashed lines), which hosts a set of low-energy states. (b) The local density of states (LDOS) distribution $\rho$ for the lowest-energy ABS inside the vortex.  The non-uniform chemical potential confines the ABS close to the upper surface.}
\label{fig: 1}
\end{figure}

Here we aim to answer this question. We show that, by using a simple theoretical model of a 3D TI proximitized with an $s$-wave SC, the zero-energy states on the Fe(Te,Se) surface could be trivial ABSs arising as a result of the  inhomogeneity in the chemical potential or, equivalently, in the electrostatic potential. Such inhomogeneity could find their origin in non-uniform gating of the sample, in the presence of impurities on the surface or in the bulk, as well as in  band bending near the surface  [see Fig. \ref{fig: 1}(a)]. All these inhomogeneities could be responsible for the appearance of zero-energy  ABSs [see Fig. \ref{fig: 1}(b)]. Our numerical simulations  identify regimes where a pair of MBSs located on opposite surfaces of Fe(Te,Se) is replaced by  zero-energy quasi-MBSs or ABSs localized only on the, say, top surface. A further numerical simulation of scanning tunneling spectroscopy (STS) measurements on the surface spectrum with  band bending gives qualitatively the same results as the experiments, which unveils the impossibility of experimentally distinguishing MBSs from trivial ABSs. Our results provide a new view of point on recent experiments in Fe(Te,Se) and could help to guide follow-up experiments uniquely identifying MBSs.

\emph{Model.}
We here adopt a 3D TI-SC lattice model \cite{RuiXingZhang, Ghaemi, FuChunZhang, JiangpingHu}, which has been proven to capture basic properties of superconductivity in Fe(Te,Se). The Hamiltonian of the 3D TI \cite{RuiXingZhang} in momentum space can be written as: 
$\hat{H}=\int d{\bm k}\ \hat{c}^{\dagger}({\bm k})H_0({\bm k})\hat{c}({\bm k})$
with $\hat{c}({\bm k})=[\hat{c}_{1 \uparrow}({\bm k}), \hat{c}_{1\downarrow}({\bm k}), \hat{c}_{2\uparrow}({\bm k}), \hat{c}_{2\downarrow}({\bm k})]^T$, where the vector consists of annihilation operators acting on electrons characterized by the orbital and spin degrees of freedom. Here,
$H_0({\bm k})=\hbar v\sigma_x(s_x \sin{k_x a}  +s_y \sin{k_y a} + s_z \sin{k_z a})+m({\bm k})\sigma_z$ with $m({\bm k})=m_0-m_1(\cos{k_x a}+\cos{k_y a})-m_2\cos{k_z a}$. The Pauli matrices $\sigma_i$ and $s_i$ act on orbital and spin space, respectively. In our lattice model we set $\hbar v=10\ {\rm meV}$ and $a=2.5\ {\rm nm}$ (the lattice constant of the cubic lattice to be used as the length unit). 

We work with an isotropic $s$-wave superconducting pairing, which is sufficient to investigate the excitations within a single vortex line threading opposite surfaces in Fe(Te,Se) SC. The Bogoliubov-de-Gennes (BdG) Hamiltonian describing a uniform 3D TI-SC setup can be written as: 
\begin{eqnarray}
H_{\rm BdG}({\bm k})=
\left(
\begin{array}{cc}
H_0({\bm k})-\mu_0 & i\Delta_0 s_y\\
-i\Delta_0 s_y & \mu_0-H_0^*(-{\bm k})
\end{array}
\right)
\label{eq: 1}
\end{eqnarray}
in the basis $[\hat{c}({\bm k}), \hat{c}^{\dagger}(-{\bm k})]$. Here $\mu_0$ is the chemical potential and $\Delta_0$ represents the superconducting gap. 
 By choosing the parameters: $m_0=-4$,  $m_1=-2$, and $m_2=1$ (in units of $\hbar v$), $H_0({\bm k})$ describes a strong TI with helical surface states and a band gap $\Delta_{\rm TI}=10\ {\rm meV}$ at the band inversion point  $Z$. To fit to the regime of experiments \cite{HongJunGao1, Shin, HongDing1, Tamegai, HongJunGao2}, we set the SC gap $\Delta_0=1.8\ {\rm meV}$.  

It is predicted that, inside the vortex of the 3D TI-SC, a pair of MBSs  appears at the ends of the vortex line \cite{Kane1}. Here, we consider a single vortex line induced by an external magnetic field. The superconducting pairing potential increases linearly with $r$ inside the vortex and is constant outside of the boundary $r_0$: $\Delta({\bm r})=\Delta_0[\Theta(r-r_0)+\Theta(r_0-r)\cdot r/r_0]e^{i\theta}$. The coordinates in the cylindrical system are $(r, \theta)$, and $\Theta (r)$ is the step function. The orbital and Zeeman effects of the magnetic field can be neglected in our model due to the large London penetrating depth in Fe(Te,Se) SC \cite{Prozorov}. By considering a finite-size system along the $z$ axis, our model well reproduces a pair of localized MBSs on the surfaces [see Supplementary Material (SM)~\cite{Supplementary}].

\emph{Non-uniform potential close to the surface and ABSs.}
In view of possible inhomogeneities in Fe(Te,Se), we use a position-dependent electrostatic potential $U(z)$. Although the chemical potential $\mu_0=5\ {\rm meV}$ on the surface is inside the TI gap, the inhomogeneity in the potential due to gating or the presence of impurities might result in the local chemical potential $\mu(z)\equiv \mu_0+U(z)$ (measured from the Dirac point of the TI) to cross conduction or valence bands. We note that the electrochemical potential stays constant and there is no current. For simplicity, we first consider a linear potential $U(z)=\kappa \hbar v(N_z-z)/(N_z-1)$, where $N_z$ is the height of the sample. The strength of the electric field is characterized by the parameter $\kappa$. Different types of $U(z)$ profile are considered in the SM \cite{Supplementary}. We find that the obtained results do not depend on a particular shape of the potential.  It is also not necessary to have the gradient over the entire sample. It is sufficient that $\mu$ is non-uniform on the surface.

\begin{figure}
\centering
\includegraphics[width=8.7cm, clip=]{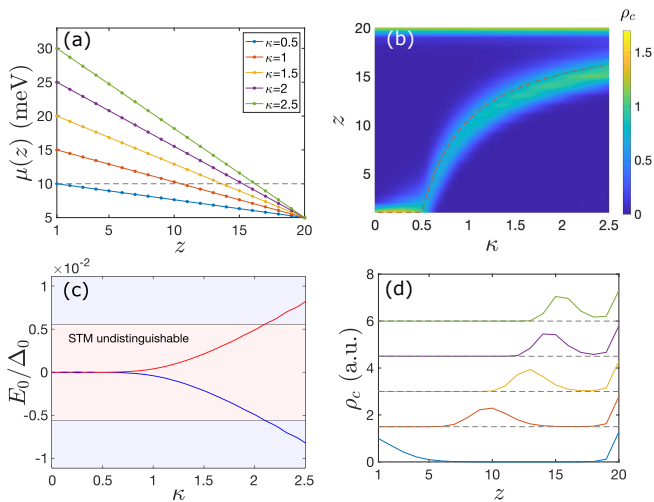}
\caption{
(a) A set of chemical potential profiles $\mu(z)$ (relative to the Dirac point of 3D TI) with different strength of electric field $\kappa$. The dashed line shows the critical chemical potential $\mu_c\approx 10\ {\rm meV}$ where the VPT happens. (b) The LDOS distribution $\rho_c$ along the center vortex line of the lowest energy state $E_0$ under varying $\kappa$. (c) The energy $E_0$ of the lowest bound state as a function of $\kappa$.  For larger values of $\kappa$, two MBSs overlap and merge into a single ABS.  In the STM undistinguishable energy range (pink color), one cannot distinguish between two types of bound states. (d) A set of LDOS curves with different $\kappa$, showing the crossover from MBSs to ABSs. Each curve is shifted with an equal distance of 1.5 (see the dashed grid). Here (a) and (d) share the same legend. The radius of the vortex is fixed to $r_0=5$. The system size is $N_x=N_y=31$, and $N_z=20$ and the surface chemical potential is $\mu_0=5\ {\rm meV}$.}
\label{fig: 2}
\end{figure}

In Fig. \ref{fig: 2}(a) we plot a set of profiles of $\mu(z)$ by varying the gradient $\kappa$. A vortex phase transition (VPT), where the line vortex changes from being a 1D topological SC to a normal SC \cite{Vishwanath, Hughes, Rossi}, occurs if  $\mu(z)$ reaches a critical value $\mu_c$ [see dashed line in Fig. \ref{fig: 2}(a)]. To calculate $\mu_c$ one may calculate the $\pi$-Berry-phase point by setting $k_z=\pi$ in our model \cite{Vishwanath, Supplementary}, or numerically find the gap closing point of the lowest energy bands by considering a periodic system along $z$-direction.
The point at which $\mu(z)$ reaches $\mu_c$  determines the spatial position at which the VPT occurs and the bottom MBS resides. When the gradient $\kappa$ is small ($\kappa <0.5$), one can observe a well-localized MBS on the top and one at the bottom surface, respectively. As $\kappa$ is increased, the bottom of the sample is in the topologically trivial phase and
one expects that the MBS on the bottom surface moves upward along with the boundary of VPT and gets closer with the top MBS until they start to overlap and merge into an ABS or, alternatively, into two quasi-MBSs. 

A numerical calculation on the local density of states (LDOS) $\rho_c$ along the center of the vortex line confirms our expectations, as can be seen in Fig. \ref{fig: 2}(b). An estimated VPT boundary $z_c$ could be found  by solving the equation $\mu_0+U(z_c)=\mu_c$,
also shown by a dashed line for comparison.  A crossover from MBSs to ABSs occurs as $\kappa$ gets larger. The energy splitting is $E_0=0.76\ \mu {\rm eV}$ for the gradient $\kappa=1$. This corresponds to a weak electric field: $\epsilon=-\partial U(z)/ \partial z\approx 0.21\ {\rm meV/nm}$ to be generated easily by some point defects or disorder \cite{Hou1, Stroscio1, Stroscio2}. In Fig. \ref{fig: 2}(d) we plot the profiles of the LDOS along the vortex line with different $\kappa$. As $\kappa>1$, the overlap of the wave functions becomes large, and these states can be identified as ABSs. Experimentally the resolution of the STM tip is about 20 $\mu {\rm eV}$ \cite{Tamegai}. In Fig. \ref{fig: 2}(c) we label the energy range which cannot be distinguished by STM by pink color. Within this region, one cannot distinguish MBSs with ABSs. Only when $\kappa$ increases to about 2.1 can this energy splitting be measured with a manifestation of double peaks around zero in STS.

\begin{figure}
\centering
\includegraphics[width=8.7cm, clip=]{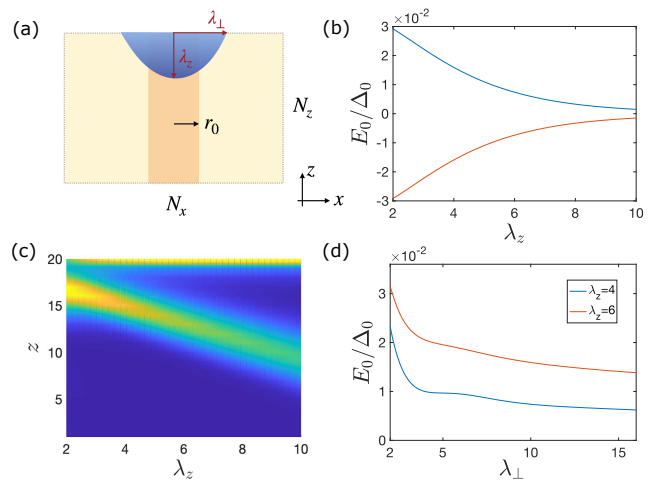}
\caption{
(a) The impurity potential $U_{\rm im}({\bf r})$ (blue) in the $xz$-plane of the 3D TI-SC. 
The lowest energy $E_0$ as a function of (b) $\lambda_z$ (for $\lambda_\perp=10$) and (d) $\lambda_\perp$. Even a short range potential of impurity creates a nearly zero-energy bound state, which could be confused with MBSs.
(c) For small values of $\lambda_z$, the ABS is localized at the top surface, as can be seen in LDOS (for $\lambda_\perp=10$). Here we set $z_0=N_z$, $U_0=8{\rm meV}$, and the chemical potential far away from the impurity $\mu_0=13\ {\rm meV}$ corresponds to the trivial phase. Other parameters are the same with Fig. 2.}
\label{fig: 3}
\end{figure}

\emph{A point impurity on the surface.}
The surface of Fe(Te,Se) usually contains impurities or defects whose spatial dimensions are comparable to  the size of the vortex \cite{HongJunGao1, Shin, HongDing1, Tamegai, HongJunGao2}, which may also influence the observation of the low-energy states. To include this effect, we consider a single impurity close to the top surface with $U_{\rm im}({\bf r})=-U_{0}e^{-R({\bf r})}$, where $R\equiv \sqrt{(x^2+y^2)/\lambda^2_ {\perp}+(z-z_0)^2/\lambda^2_z}$ with  $\lambda_{\perp}$ ($\lambda_z$) being the in-plane (out-of-plane) characteristic length scale, see Fig. \ref{fig: 3}(a). 
By changing $\lambda_{\perp}$ or $\lambda_z$, one can control the size of an area in which $\mu({\bf r})<\mu_c$. As a result, impurities close to the vortex center can generate nearly zero-energy ABSs, see Fig. \ref{fig: 3}. Even if the impurity potential is short-ranged, the ABS has almost zero energy. In this regime, it will be rather difficult to distinguish such local ABSs from MBSs localized both on the top and bottom TI surfaces. Generally, the smaller $\lambda_z$ is, the more localized is the zero-energy ABS next to the top surface, see Fig. \ref{fig: 3}(c). 
If one varies the width of the potential $\lambda_{\perp}$, the spatial positions of the ABSs along the $z$ axis stays unchanged \cite{Supplementary}, while their energy gets closer to zero for larger $\lambda_{\perp}$, mimicking MBSs.

\begin{figure}
\centering
\includegraphics[width=8.7cm, clip=]{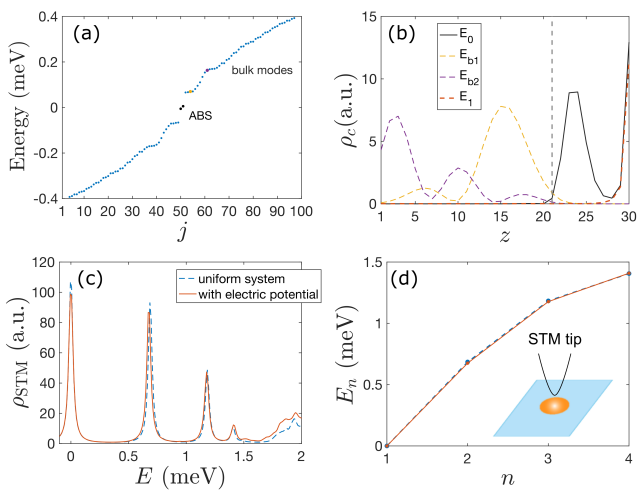}
\caption{
(a) The low-energy spectrum contains ABSs localized at the surface. An electrostatic potential is modeled by $U(z)=\kappa \hbar v[\Theta(21-z)+\Theta(z-21)(N_z-z)/(N_z-21)]$ with $\kappa=0.8$ and $N_z=30$.
(b) The LDOS $\rho_c$ along the center vortex line for the ABSs localized at the top surface ($E_0=5.6\ \mu{\rm eV}$ and  $E_1=0.676\ {\rm meV}$) and for the two bulk states ($E_{b1}=70\ {\mu}{\rm eV}$ and $E_{b2}=163 {\mu}{\rm eV}$).
The vertical dashed line denotes the boundary between the uniform and non-uniform regions at $z=21$. 
The 3D version of the ABS LDOS is also shown in Fig. \ref{fig: 1}. (c) The red (blue) line shows the results of the simulated LDOS $\rho_{\rm STM}$ of the STS measurement on the surface of Fe(Te,Se) for a system with a non-uniform (uniform) electrostatic potential hosting zero-energy ABSs (MBSs). The peaks are originating from ABSs with energy $E_n$ localized at the vortex.
(d) The energies $E_n$ extracted from the LDOS curves are equidistantly distributed in both cases. The other parameters are the same as in Fig.~2. }
 \label{fig: 4}
\end{figure}

\emph{Simulation of STM measurements on the surface of Fe(Te,Se) SC.}
Next, we study how trivial nearly zero-energy ABSs can be detected in STM experiments. To simulate realistic experimental samples we consider a thick layer with a non-uniform electrostatic potential. 
The lowest-energy ABS localized at the top surface inside the vortex is very close to zero energy ($E_0 =5.6\ \mu{\rm eV}$), see  Fig. \ref{fig: 4}(a). Its LDOS (3D LDOS) distribution along the vortex line is shown with black curve in Fig.~\ref{fig: 4}(b) [in Fig.~\ref{fig: 1}(b)]. The next ABS  also localized on the surface ($E_1=0.676\ {\rm meV}$) is indicated by the red dashed line.  Compared with clean systems (see Supplementary Material~\cite{Supplementary}), in addition to these ABSs localized  next to the impurity,  there are many low-energy bulk states localized in the vortex. In Fig. \ref{fig: 4}(b),  for comparison, we also plot the LDOS distribution of two bulk states ($E_{b1}=70\ {\mu}{\rm eV}$ and $E_{b2}=163\ {\mu}{\rm eV}$) along the vortex line.  We note that the ABSs  and low-energy bulk states are well separated spatially. Although such bulk states can slightly leak into the region with non-uniform potential, they never reach the surface
and, thus, cannot be detected by STS measurements at the surface.

Next, we calculate numerically the LDOS inside the vortex on the top surface to simulate STS measurements. The LDOS at energy $E$ is defined as \cite{Hou2}:
\begin{eqnarray}
\rho_{\rm STM}(E)=- {\rm Im}\sum_{|{\bm r}_i-{\bm r}_c|<r_0} {\rm Tr}\,{ G}^r({\bm r}_i,E)/\pi,
\label{eq: 3}
\end{eqnarray}
where ${ G}^r({\bm r_i},E)=[(E+i\eta)-{ H}_{\rm BdG}]^{-1}$ is the retarded Green function. Here, ${\bm r}_c$ is the position of the vortex center and the summation in Eq. (\ref{eq: 3}) is over the vortex region on the top surface. An imaginary term with $\eta=0.02\ {\rm meV}$ is included to account for the energy broadening in the STS measurement. 

The localized ABSs are responsible for a series of well-sequenced LDOS peaks with a pronounced zero-energy peak, see Fig. \ref{fig: 4}(c) [solid curve]. As expected, the low-energy bulk states cannot be read out in the LDOS curve. To compare with signals coming from MBSs, we also show the LDOS curve for a case of uniform electrostatic potential in the topological phase (see the dashed curve). Only little difference can be seen between these two LDOS curves, indicating that normal measurements on the surface fail to distinguish ABSs with MBSs. In Fig. \ref{fig: 4}(d),  we extract these ABS energy levels $E_n$ and plot them as a function of the level index $n$. It is different from the half-integer sequence $E_n \propto n+1/2$ of excitations in vortices of normal SC \cite{Volovik}. Instead, the relation $E_n \propto n$ is reproduced typical for topological vortices is reproduced, see Fig. \ref{fig: 4}(d). This result is in remarkable agreement with  recent experimental observations interpreted as MBSs~\cite{HongDing1}, but, here, it originates from trivial ABSs.

\emph{Discussion and conclusions.}
In conclusion, we have clarified the ambiguity of recent experimental observations on the zero-energy states in iron-based SC Fe(Te,Se), using a simple model of 3D TI proximity with an $s$-wave SC. We show that, if there are non-uniform electrostatic potentials near the surface of Fe(Te,Se) as a result of chemical inhomogeneity, presence of impurities, applied bias, these zero-energy states might well be ABSs instead of MBSs. Numerical simulations show that the ABSs have energies close to zero and are localized close to the top surface in contrast to MBSs assumed to be localized on both top and the bottom surfaces. These ABSs are thus incapable for topological quantum computing as they are easily destroyed by decoherence or impurity scattering. Still, these ABSs show other features usually associated with MBSs, unveiling the failure of STS measurements on distinguishing them. All these results imply that  more careful studies are needed to ensure beyond reasonable doubt that Fe(Te,Se) SCs provide a  robust platform for topological quantum computing.

Our results indicate again that it is far from sufficient to confirm MBSs by solely measuring the ZBP peaks. A rigorous characterization of the wavefunction, e.g. its localization, is also a necessary condition in searching for this exotic states. 
To make progress in this field, it seems indicated to go beyond static detection methods and instead use dynamic schemes, such as non-Abelian braiding operations on MBSs \cite{Fisher, Beenakker3}, or other exotic properties in quantum transport concerned with them \cite{Kane3, Tewari2, Vishveshwara, Bruder, YingTaoZhang}. 

\begin{acknowledgements}
{\it{Acknowledgements.}}
We thank Vardan Kaladzhyan, Henry Legg, and Daniel Loss for helpful discussions. This work was supported by the Swiss National Science Foundation (SNSF) and NCCR QSIT. This project received funding from the European Union's Horizon 2020 research and innovation program (ERC Starting Grant, grant agreement No 757725).
\end{acknowledgements}

\end{document}


\def\theequation{S\arabic{equation}}

\title{Supplementary Material for ``Zero-energy Andreev bound states in iron-based superconductor Fe(Te,Se)"}

\author{Zhe Hou}
\affiliation{Department of Physics, University of Basel, Klingelbergstrasse 82, CH-4056 Basel, Switzerland}

\author{Jelena Klinovaja}
\affiliation{Department of Physics, University of Basel, Klingelbergstrasse 82, 
		CH-4056 Basel, Switzerland}

\maketitle
\section{Details on simulation of finite-size systems} 

\begin{figure}
\centering
\includegraphics[width=9cm, clip=]{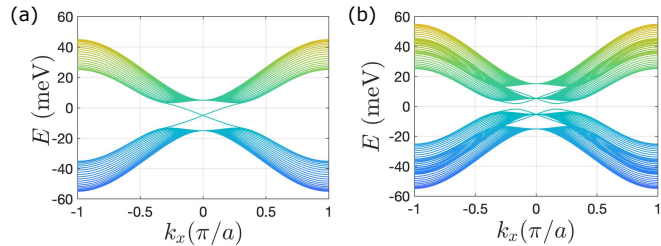}
\caption{(Color online)
Energy spectrum (a) of 3D TI and (b) of 3D TI-SC without vortices. The number of layers is set $N_z =20$ and the momentum $k_y=0$. The chemical potential $\mu_0=5\ {\rm meV}$ and the SC gap $\Delta_0=1.8\ {\rm meV}$.}
\label{fig: S1}
\end{figure}

In the main text, we have used a 3D TI coupled with an $s$-wave SC to simulate the experimental sample Fe(Te,Se). Without the vortex, the 3D TI-SC respects the time-reversal symmetry and belongs to the normal SC phase with no topological surface or boundary states. In Fig. S\ref{fig: S1}(a), we first plot the energy spectrum of a TI slab without SC pairing term. The TI slab has a finite layer number $N_z=20$ and hosts topological surface states within the band gap (here the momentum $k_y$ is set to zero). The chemical potential $\mu_0$ is chosen to be 5 meV, so the Dirac point is shifted below the zero energy. The band gap of the TI at $(k_x=0, k_y=0)$, $\Delta_{\rm TI}$, is around 10 meV, and we use it to estimate the localization length of the surface states by the relation: $\xi_z=\hbar v_F/(\Delta_{\rm TI}-\mu_0)$, with $v_F=va$ being the Fermi velocity. With a uniform SC pairing term $\Delta_0=1.8\ {\rm meV}$, the states at the Fermi surface are gapped and the system is in the normal SC phase. Further introducing a vortex line along $z$-direction through the SC, a pair of MBSs would appear  inside the vortex on the top and bottom surfaces of the SC, as stated by Fu and Kane \cite{Kane1}. Since these MBSs are originating from the TI helical surface states they share the same decaying length $\xi_z$ into the bulk.\\

Experimental samples of Fe(Te,Se) can be regarded to be infinitely large. Simulating the MBSs in such a huge system is numerically impossible. However, since we only investigate the excitation properties within one vortex line, a finite-size system with moderate sizes should be enough to simulate the Fu-Kane model. The main effect of the finite-size is enhancing the coupling between the MBSs on the top and bottom surfaces so that their energies split from zero. This coupling can be realized in two regimes: 1) The wavefunctions of the MBSs spread over the side surfaces and couple on the surfaces, similar to the coupling in 2D system \cite{Biswas, MengCheng}. This coherence length on the surfaces is determined by the parameter $\xi=\hbar v_F/\Delta_0$, which is about $6a$ or 15 nm within our parameter regime. 2) The MBSs penetrate deeply into the bulk and couple in $z$-direction. Here, a characteristic lengthscale is  $\xi_z=2a$. Both couplings decay as the system size increases.  \\

To simulate the MBSs within the vortex of a 3D TI-SC, we first write the tight-binding (i.e. discretized) model of the 3D system:
\begin{eqnarray}
\hat{H}_{BdG}=\sum_i \left[ f^\dagger_i t_{0i} f_i+ \left( f^\dagger_i t_x f_{i+\hat{e}_x}+f^\dagger_i t_y f_{i+\hat{e}_y}
+f^\dagger_i t_z f_{i+\hat{e}_z}+H.c. \right) \right],
\label{eq: S1}
\end{eqnarray}
where $f^\dagger_{i}$ is the creation operator of the Bogoliubov quasiparticle at site ${\bm{r}_i}$. In addition, the vector $f^\dagger_{i}$ can be rewritten as $f^\dagger_i=\left[ \hat{d}^\dagger_{1 \uparrow}(i),\hat{d}^\dagger_{1\downarrow}(i),\hat{d}^\dagger_{2\uparrow}(i),\hat{d}^\dagger_{2\downarrow}(i),\hat{d}_{1\uparrow}(i),\hat{d}_{1\downarrow}(i),\hat{d}_{2\uparrow}(i),\hat{d}_{2\downarrow}(i) \right]$, with the creation operator $\hat{d}^\dagger_{l,\alpha}(i)$ with orbital $l=1,2$ and spin $\alpha=\uparrow,\downarrow$ defined as:
\begin{eqnarray}
\hat{d}^\dagger_{l,\alpha}(i)=\frac{1}{\sqrt{N}}\sum_i e^{-i {\bm k}\cdot {\bm r}_i}\hat{c}^\dagger_{l,\alpha}({\bm k}),
\label{eq: S2}
\end{eqnarray}
with $N$ the total number of the lattice sites, and $\hat{c}^\dagger_{l,\alpha}({\bm k})$ is the creation operator defined in the main text. The on-site and hopping terms in Eq. (\ref{eq: S1}) are:
\begin{align}
t_{0i}&=\left[
\begin{matrix}
m_0 \sigma_z\otimes s_0-\mu_0 & i \Delta(\bm{r}_i)\sigma_0\otimes s_y\\
-i\Delta(\bm{r}_i) \sigma_0\otimes s_y & \mu_0-m_0\sigma_z\otimes s_0
\end{matrix}
\right], \nonumber \\
t_x&=\left[
\begin{matrix}
\frac{\hbar v}{2i}\sigma_x \otimes s_x -\frac{m_1}{2}\sigma_z\otimes s_0 & 0\\
0 & \frac{\hbar v}{2i}\sigma_x \otimes s_x+\frac{m_1}{2}\sigma_z\otimes s_0
\end{matrix}
\right], \nonumber \\
t_y&=\left[
\begin{matrix}
\frac{\hbar v}{2i} \sigma_x \otimes s_y-\frac{m_1}{2}\sigma_z\otimes s_0 & 0\\
0 & -\frac{\hbar v}{2i}\sigma_x \otimes s_y+\frac{m_1}{2}\sigma_z\otimes s_0
\end{matrix}
\right], \nonumber \\
t_z&=\left[
\begin{matrix}
\frac{\hbar v}{2i}\sigma_x \otimes s_z -\frac{m_2}{2}\sigma_z\otimes s_0 & 0\\
0 & \frac{\hbar v}{2i}\sigma_x \otimes s_z +\frac{m_2}{2}\sigma_z\otimes s_0
\end{matrix}
\right].
\label{eq: S3}
\end{align}
Here $\Delta({\bm r}_i)$ is the pairing potential describing the SC vortex, as defined in the main text. By diagonalizing 
the Hamiltonian in Eq. (\ref{eq: S1}), we can get the low-energy excitation states including the MBSs inside the SC gap $\Delta_0$.\\

\begin{figure}
\centering
\includegraphics[width=9cm, clip=]{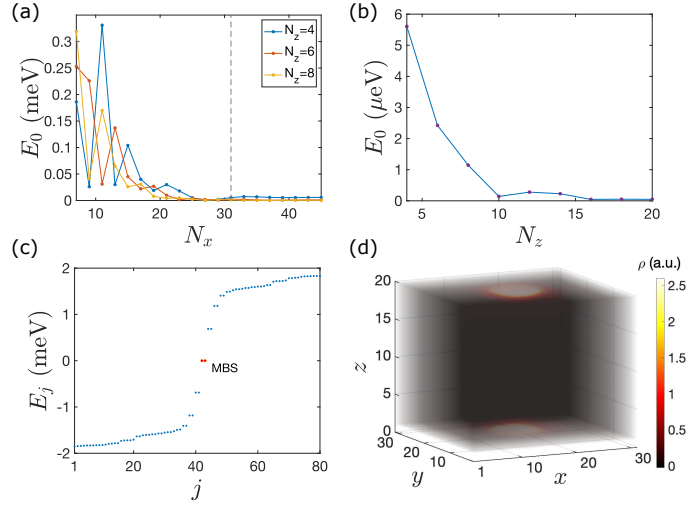}
\caption{(Color online)
Scaling of the MBS energy $E_0$ in a finite-size system. (a) The lowest energy $E_0$ decays after exhibiting  several oscillations as one changes the width $N_x$ of 3D TI-SC. The number of layers  $N_z$ is fixed to be 4, 6, and 8. (b) The lowest energy $E_0$ shown as a function of $N_z$. Here, the width is fixed to be $N_x=N_y=31$. (c) The low-energy spectrum inside the vortex of a 3D TI-SC of finite-size geometry defined  by $N_x=N_y=31$ and $N_z=20$. The red dots represent the MBSs. (d) The 3D LDOS distribution of the MBSs from the panel (c). Here, the chemical potential is uniform, $\mu_0=5\ {\rm meV}$. The radius of the vortex is $r_0=5$.}
\label{fig: S2}
\end{figure}

Next, we consider a finite-size system, $N_x=N_y$, and plot the energy $E_0$ of the lowest state as a function of $N_x$ and $N_z$ in Fig. S\ref{fig: S2}(a) and Fig. S\ref{fig: S2}(b), respectively. An exponential decay along with sinusoidal oscillations is observed as we vary $N_x$. These oscillations vanish when $N_x=31$, where $E_0$ approaches a stable value that is dominated by the coupling along the  $z$-direction. Then we set $N_x=31$ and study the scaling behaviour with the number of layers $N_z$. A quicker decay is observed due to the shorter coherence length $\xi_z$ in $z$ direction. A small increase of $E_0$ can be observed at $N_z=10$, but $E_0$ decays into a small energy range in the order of 0.01 $\mu{\rm eV}$ after $N_z=16$. Compared with the energy range we investigated in the main text, this small energy splitting can be ignored and $E_0$ can be regarded as a zero energy in our finite-size system.\\

In our calculations to reproduce results of the Fu-Kane model, we choose $N_x=N_y=31$ and $N_z=20$, which corresponds to a $\rm 75 \times 75 \times 50\ nm^3$ system. A pair of well-separated MBSs are shown in Fig. S\ref{fig: S2}(c), with their LDOS distribution shown in Fig. S\ref{fig: S2}(d). In addition, the Caroli-de Gennes-Matricon bound states (CBSs) with higher energies are given by 0.68, 1.18, and 1.40 meV, respectively, which is consistent with experimental observations~\cite{HongDing1} of 0.65, 1.37 and 1.93 meV.\\

\section{Determining the critical chemical potential $\mu_c$ of the VPT} 

\begin{figure}[b]
\centering
\includegraphics[width=9cm, clip=]{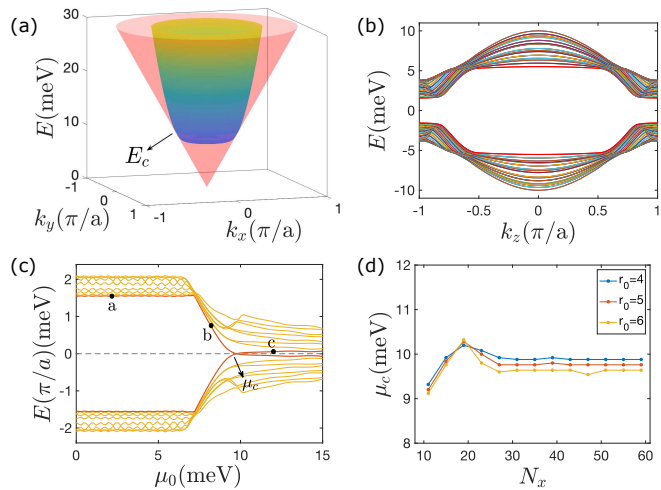}
\caption{(Color online) (a) The conduction band of the low-energy Hamiltonian in Eq. (\ref{eq: S4}) as shown with the blue-yellow color. Here, the band with the dispersion $\hbar v_F |k_{\perp}|$ of the massless Dirac fermion is also shown as a comparison (see the pink band). At the critical energy $E_c$ the two bands are tangent and the particle circling around the Fermi surface of the 3D TI acquires a $\pm \pi$ Berry phase. (b) The band structure of a periodical quasi-1D vortex system with chemical potential $\mu_0=5 {\rm meV}$. Here the lowest 100 energies per $k_z$ are given. (c) The low-energy bands at $k_z=\pi/a$ as functions of the chemical potential $\mu_0$. A gap closing is observed at the critical chemical potential $\mu_c$, which implies a VPT or a topological phase transition. In (b) and (c), the width of the system is set to $N_x=N_y=31$, and the radius of the vortex is $r_0=5$. (d) Dependence of the critical chemical potential $\mu_c$ on the system parameters $N_x$ and $r_0$. Here we have fixed $N_y=N_x$. In (b,c,d), the SC gap is set to $\Delta_0=1.8\ {\rm meV}$.}
\label{fig: S3}
\end{figure}

In understanding the advent of MBSs in the SC vortex of 3D TI-SCs, one may regard the surface SC as a spinless $p_x+ip_y$ SC, which harbours MBSs inside the vortex, or regard the vortex line as a 1D topological SC which hosts Majorana bound states at its ends, similar to the Rashba nanowires in proximity with SC \cite{Kat}. The second way in understanding the MBSs is more convenient when we analyze the vortex phase transition (VPT) \cite{Hosur}. The VPT here is a topological phase transition where the 1D vortex system changes from a topological SC into a trivial SC. The governing parameter is the chemical potential $\mu$, and there exists a critical chemical potential $\mu_c$ at which the VPT occurs. To determine the critical chemical potential $\mu_c$, we consider periodic boundary conditions of the vortex system in $z$-direction, i.e. we choose $k_z$ as a good quantum number, and find the gap closing point of the band dispersion $E(k_z)$. The translation symmetries along $x$ and $y$ directions are both broken due to the presence of the SC vortex. Here we consider that the size of the vortex is small compared with the cross-section in the $xy$-plane, so the vortex bound states are all generated from the bulk states of the 3D TI. The gap closing point of $E(k_z)$ indicates the zero-energy states or solutions inside the vortex system. As superconductivity is dominated by the electrons around the Fermi surface (FS), we first analyze the FS of the 3D TI without SC.  As pointed out by Hosur et al.~\cite{Hosur} and  by Volovik~\cite{Volovik}, the $\pi$ Berry phase of electrons acquired around the FS in the $k_xk_y$-plane would induce a 1/2 phase shift in the Bohr-Sommerfield quantization rule, and thus yields an exactly zero-energy solution inside the SC vortex. The Berry phase criteria here is independent of the SC parameters and is only determined by the FS property \cite{Hosur}. To find this point, we note that the TI Hamiltonian $H_0({\bm k})=\hbar v \sigma_x (s_x \sin{k_x a} +s_y \sin{k_y a}+ s_z \sin{k_z a})+m({\bm k}) \sigma_z$, with $m({\bm k})=m_0-m_1(\cos{k_x a}+\cos{k_y a})-m_2 \cos{k_z a}$, can mimic a Dirac spectrum around the eight time-reversal invariant points $(k_x, k_y, k_z)$ with $k_i=0, \pi/a$ $(i=x,y,z)$. Since the momentum $\hbar k_z$ is a good quantum number, we effectively  deal with a 2D system. As we have known that the Berry phase acquired for a 2D massless Dirac fermion is exactly $\pi$ when the Dirac fermion makes a closed loop in the momentum space \cite{Hosur}, the most convenient way to find the FS with $\pi$ Berry phase of the 3D TI is to mimic its Hamiltonian to a 2D massless Dirac Hamiltonian at finite energy. This analogy can be implemented by setting $\sin{k_z a}$ to zero.  So immediately we know that $k_z$ can only take the two values: zero and $\pi/a$. Then, since we have chosen the parameters $m_0=-4, m_1=-2$ and $m_2=1$ (in the unit of $\hbar v$) in our model, we work with low-energy solutions if $k_z=\pi/a$.  Then, we get $m({\bm k})=-3+2(\cos{k_x a}+\cos{k_y a})$ and the low energy states in the spectrum are around $(k_x, k_y)=(0,0)$. So next, we expand the Hamiltonian $H_0({\bm k})$ in $k_x$ and $k_y$ around $(0,0,\pi/a)$, i.e. the band inversion point $Z$: 
\begin{eqnarray}
H_0({\bm k})\approx 
\hbar v_F\sigma_x{\bm s}\cdot {\bm k_{\perp}}+\sigma_z(m_0-2m_1+m_2+m_1 k_{\perp}^2a^2/2);
\label{eq: S4}
\end{eqnarray}
where ${\bm k_{\perp}}=(k_x, k_y)$ and $k_{\perp}=\sqrt{k_x^2+k_y^2}$. The dispersion of $H_0({\bm k})$ is then:
\begin{eqnarray}
E({\bm k_{\perp}})=\pm \sqrt{ \left( m_0-2m_1+m_2+m_1k_{\perp}^2 a^2/2 \right) ^2+\hbar^2 v^2_Fk_{\perp}^2}.
\label{eq: S5}
\end{eqnarray}
Each energy branch is doubly-degenerate. In Fig. S\ref{fig: S3}(a) we plot the positive branch of the band (see the blue-yellow colored band). The bottom of the band shows a Mexican-hat shape with a minimum energy $E_{min}=\frac{\hbar v_F}{-m_1a^2}\sqrt{-2m_1 (m_0-2m_1+m_2) a^2-\hbar^2v^2_F}\approx 8.7 {\rm meV}$ at $k_{\perp}=\sqrt{\frac{2(m_0-2m_1+m_2)+\hbar^2 v^2_F}{-m_1 a^2}}$. This energy defines the gap of the bulk 3D TI. In Eq. (\ref{eq: S4}) if the mass term $m({\bm k})$ is zero, i.e. $m_0-2m_1+m_2+m_1 k_{\perp}^2a^2/2=0$, the Hamiltonian can be decoupled into a pair of massless Dirac equations. To see this visually, we plot the band of a massless Dirac fermion with dispersion $E=\pm \hbar v_F |k_{\perp}|$ as a comparison [see the pink band in Fig. S\ref{fig: S3}(a)]. When $k_{\perp}=k_c=\frac{1}{a}\sqrt{2(2m_1-m_0-m_2)/m_1}$, the 3D TI band is tangent to the massless Dirac band at energy $E_c=\hbar v_F k_c=10\ {\rm meV}$. If the particle makes an enclosed loop at the iso-energy curve $E_c$, it will acquire a $\pm \pi$ Berry phase, which is reminiscent of the $\pm \pi$ Berry phase of a massless Dirac fermion when it makes arbitrary closed motion in momentum space as long as the Dirac point is enclosed. The tangent energy, $E_c$, is defined as the critical energy here. When considering superconductivity in the 3D TI, this energy is also used to define the critical chemical potential $\mu_c$. Since the SC gap $\Delta_0$ is small compared with the energy gap of the 3D TI, the Bogoliubov quasiparticles in the SC are mainly formed from the unpaired particles of the 3D TI at the FS. By further inducing a SC vortex into the 3D TI-SC, the Bogoliubov quasiparticles would be bound inside the vortex potential and their motions would form enclosed loops in the momentum space. This enclosed motion generates a Berry phase for the Bogoliubov quasiparticles, which can be thought to be the same as the one obtained by the unpaired particles at the FS of the 3D TI due to the tiny SC pairing potential inside the vortex. So, if the chemical potential $\mu=\mu_c\equiv E_c$, the Berry phase of the Bogoliubov quasiparticle inside the vortex is exactly $\pi$, which induces a 1/2 phase shift, and thus one obtains a pair of zero-energy states inside the SC vortex according to the Bohr-Sommerfield quantization rule \cite{Hosur}. This also implies the gap closing or, equivalently, the topological phase transition inside the vortex system. Away from $E_c$, the particle in 3D TI turns into a massive one and cannot get a $\pm \pi$ Berry phase, so the zero-energy solution disappears and a gap is reopened in $E({\bm k})$.
In our system, for chemical potential $\mu<\mu_c$, the 1D vortex line belongs to the TSC phase and hosts two MBS at the ends when open boundaries are considered. Tuning the chemical potential above $\mu_c$, the VPT from 1D TSC to normal SC occurss. At VPT, the MBSs would merge into the bulk states and disappear. \\

\begin{figure}[b]
\centering
\includegraphics[width=9cm, clip=]{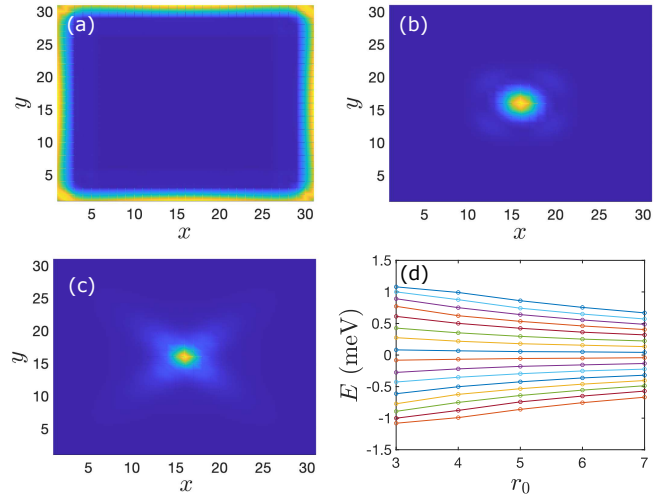}
\caption{(Color online) The wave function probability density of the lowest energy state at  the system with the following values of the chemical potential: (a) $\mu_0=2\ {\rm meV}$, (b) $\mu_0=8\ {\rm meV}$, and (c) $\mu_0=12\ {\rm meV}$. The wavefunctions here correspond to the three selected stated indicated by points $a, b, c$ in Fig. S\ref{fig: S3}(c). The radius of the SC vortex is $r_0=5$. (d) Dependence of the low-energy spectrum on the radius of the SC vortex $r_0$ for the chemical potential fixed to $\mu_0=12\ {\rm meV}$. In all the panels we set $N_y=N_x=31$, and the SC gap $\Delta_0=1.8\ {\rm meV}$.
}
\label{fig: S4}
\end{figure}

One can also obtain the critical chemical potential $\mu_c$ numerically by directly calculating the band structure of the vortex system and find the band gap closing point. In this case open boundaries should be considered in both $x$- and $y$-directions. The tight-binding Hamiltonian has already been given in Eq. (\ref{eq: S1}). Periodic boundary conditions are given in $z$-direction so one can calculate the band $E(k_z)$ of the quasi-1D system. In Fig. S\ref{fig: S3}(b) we show the energy band as a function of $k_z$ for the chemical potential $\mu_0=5\ {\rm meV}$. The band gap is located at $k_z=\pm \pi/a$.  Then we plot the energies $E$ at $k_z=\pi/a$ as functions of the chemical potential $\mu_0$ to see how the band gap changes. As we can see in Fig. S\ref{fig: S3}(c) the band gap of the 1D vortex system undergoes a closing and reopening process around a specific chemical potential, which is just the critical chemical potential $\mu_c$. In Fig. S\ref{fig: S3}(c),  $\mu_c=9.76\ {\rm meV}$, which is very close to the one obtained from the Berry phase criteria. In Fig. S\ref{fig: S3}(d) we also investigate the parameter dependence of $\mu_c$ by varying the width $N_x$ and the size of the vortex $r_0$. As we can see, the critical chemical potential shows an obvious dependence with $N_x$ when $N_x<31$ as a result of the finite-size effects. After $N_x=31$, $\mu_c$  saturates to a fixed value. The radius of the SC vortex $r_0$ is also varied here. The value of $\mu_c$ shows a slight decrease as $r_0$ increases. \\

One may notice that the reopened gap at $\mu>\mu_c$ is relatively small compared with the band gap in the topological region [see Fig. S\ref{fig: S3}(c)]. To explain this, in Fig. S\ref{fig: S4} (a-c) we plot the distributions of the wavefunctions of the low-energy states, which corresponds to the three selected points: $a \ (\mu_0=2\ {\rm meV}), 
b \ (\mu_0=8\ {\rm meV})$, and $c \ (\mu_0=12\ {\rm meV})$ in Fig. S\ref{fig: S3}(c). As one can see, the wave-function in Fig. S\ref{fig: S4}(a) is localised on the surface of the 2D cross-section since the chemical potential $\mu_0$ is placed close to the Dirac point and only the surface states of the 3D TI participate the SC pairing. When the chemical potential increases and comes closely to the minimum energy $E_{min}$ of the band, the bulk states of the 3D TI also subjected to the SC pairing and they form the ABS inside the SC vortex [see Fig. S\ref{fig: S4}(b)] with energies smaller than the SC gap $\Delta_0$. So one can observe a decreasing energy for $\mu_0$ around 7 meV and 10 meV. Further increasing the chemical potential above the band crossing point $\mu_c$, the FS crosses deeper into the bulk states of the 3D TI, and one can still observe the ABS inside the vortex, as can be seen from Fig. S\ref{fig: S4}(c). Since the wavefunctions of these ABS are well localised around the center of the vortex and the their localisation lengths are relatively small compared with the system size ($N_x=N_y=31$), their energies are mainly determined by the radius of the SC vortex $r_0$. In Fig. S\ref{fig: S4}(d) we investigate the dependence of the low-energy spectrum of the ABS on $r_0$ by fixing the chemical potential $\mu_0=12\ {\rm meV}$. Clearly one can see that the energies decay slowly as $r_0$ is increased, but never reach zero, which is a typical behaviour of the ABS confined inside constricted regions.

\section{Non-uniform surface chemical potential} 

\begin{figure}
\centering
\includegraphics[width=8.7cm, clip=]{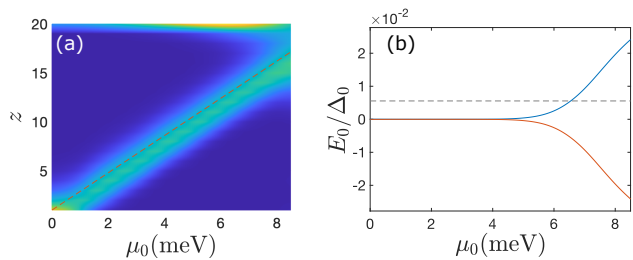}
\caption{(Color online)
Dependence of the lowest-energy state $E_0$ inside the vortex on the chemical potential at the surface $\mu_0$. (a) The LDOS distribution along the center vortex line $\rho_c$ as varying $\mu_0$. The VPT boundary $z_c$ calculated with Eq. (2) in the main text is also shown here with a dashed line. (b) The energy  $E_0$ of the lowest bound state as a function of  $\mu_0$. 
A linear part of the potential is characterized by $\kappa=1$ applied uniformly through the system. The other parameters are: $N_x=N_y=31, N_z=20$, and $r_0=5$.}
\label{fig: S5}
\end{figure}

In experimental setups, the chemical potential $\mu_0$ on the surface could  also be non-uniform with lots of domains consisting of topological and non-topological surface states \cite{HongDing1}. To capture this situation, we change the chemical potential $\mu_0$ from 0 to 8.5 meV as we consider the electric gradient characterized by $\kappa=1$. 
Basically, changing $\mu_0$ has two main effects on the MBSs. It determines the VPT boundary by the relation in Eq. (2) in the main text and it determines the localization length of the MBSs in the $z$ direction by the relation: $\xi_z=\hbar v_F/(\Delta_{\rm TI}-\mu_0)$ with $v_F=v a$ being the Fermi velocity. The second relation comes from the fact that the MZM inherits the topology of the helical surface states of TI. A larger value of the chemical potential $\mu_0$ means a shorter distance between the two MBSs [see Fig. S\ref{fig: S5}(a)] and a longer localization length of the MBSs. As a result, a sharp increase of the lowest energy $E_0$ can be observed in Fig. S\ref{fig: S5}(b) when $\mu_0>5 {\rm\ meV}$ where the two MBSs are separated by 10 layers and merge into trivial ABSs.
According to the limitation of the STS resolution, our numerical results in Fig. S\ref{fig: S5} with $\kappa=1$ give an undistinguishable ABS when  $\mu_0<7.2\ {\rm meV}$. Actually, increasing the surface chemical potential $\mu_0$ also yields a lower limitation on the surface electric fields. When the chemical potential $\mu_0$ is close to the critical chemical potential $\mu_c$, even a small perturbation on the surface would squeeze the pair of MBSs residing on the opposite surfaces into one surface and fusing them into the ABS.  

\section{Simulation of the surface potential with $\tanh(z)$ function}

\begin{figure}
\centering
\includegraphics[width=9cm, clip=]{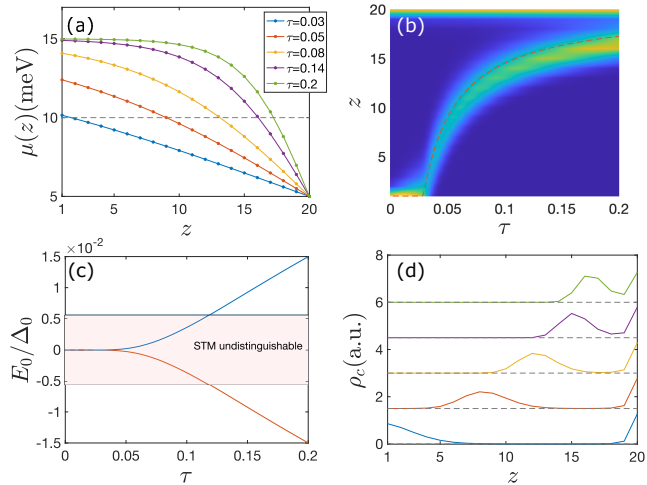}
\caption{(Color online) 
(a) A set of chemical potential profiles $\mu(z)$ with a hyperbolic surface electric potential $U(z)=U_\tau \tanh{\tau(N_z-z)}$. The dashed line shows the critical chemical potential $\mu_c$ where the VPT occurs. (b) The LDOS $\rho_c$ along the center vortex line as varying the parameter $\tau$. (c) The lowest energy $E_0$ grows as one increases $\tau$. The  energy range  indistinguishable by the STM  is indicated by the pink color. (d) The LDOS, $\rho_c$, for different values of $\tau$. The curves are shifted with an equal distance 1.5 (indicated by the dashed grid). Here (a) and (d) share the same legend. The parameters are chosen as: $N_x=N_y=31, N_z=20$, $U_\tau=10{\rm meV}$, $\mu_0=5{\rm meV}$ and $r_0=5$.}
\label{fig: S6}
\end{figure}

In the main text we considered a linear electric potential $U(z)$ near the surface and elucidate the ambiguity between MBSs and ABSs. This conclusion is also applicable to other gradually changing potentials. For concreteness here we use a hyperbolic function: $U(z)=U_{\tau} \tanh{\tau(N_z-z)}$ to simulate the surface potential. Here $U_{\tau}$ represents the amplitude of the electrostatic potential and $1/\tau$ is the characterization lengthscale of non-uniformity in the electrostatic potential. In Fig. S\ref{fig: S6}(a) we plot a set of $\mu(z)=U(z)+\mu_0$ with different $\tau$. The critical chemical potential $\mu_c$ is shown with the dashed line. Similar to the linear potential case as shown in Fig. 2 of the main text, as $\tau$ increases the MBS residing on the VPT boundary gets closer to the top MBS [Fig. S\ref{fig: S6}(b)] and their coupling increases resulting in a splitting [Fig. S\ref{fig: S6}(c)]. When $\tau>0.05$, the distance between two MBSs is less than 10 layers and the wavefunctions become delocalized [Fig. S\ref{fig: S6}(d)], and they should be called now ABSs. When the energy $E_0$ is within 10$\mu{\rm eV}$, the STS can not distinguish them from MBSs. 

\section{Discussion on the surface impurity potential}

\begin{figure}
\centering
\includegraphics[width=9cm, clip=]{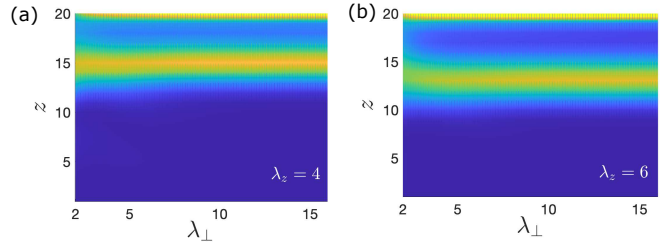}
\caption{(Color online) 
The LDOS $\rho_c$ along the center vortex line as varying the parameter $\lambda_\perp$. Here, the depth of the impurity potential is (a) $\lambda_z=4$  and (b) $\lambda_z=6$. Here, we set $z_0=N_z$, $U_0=8\ {\rm meV}$, and $\mu_0=13\ {\rm meV}$. Other parameters are the same as in Fig. 2 in the main text.
}
\label{fig: S7}
\end{figure}

In the main text we use a point impurity absorbed on the top surface of Fe(Te,Se) to simulate the unavoidable disorder or impurities in Fe(Te,Se). The impurity is responsible for a modification of the surface chemical potential: $\mu({\bf r})=\mu_0+U_{\rm im}({\bf r})$, with $\mu_0$ being the chemical potential far away from the impurity, and $U_{\rm im}({\bf r})=-U_0 e^{-R({\bf r})}$ is the impurity potential. With a vortex line threading the impurity one can expect two MBSs locating on the top surface of Fe(Te,Se) and the lower boundary of the impurity, respectively. The size of the impurity is determined by the depth $\lambda_z$ and the width $\lambda_\perp$. In Fig. S\ref{fig: S7}, we plot the LDOS along the vortex line center $\rho_c$ as varying the width $\lambda_\perp$, which corresponds to Fig. 3(d) in the main text. As we can see, the positions of the two MBSs are almost irrelevant to $\lambda_\perp$. However, the broadening of LDOS for the lower MBS is affected by $\lambda_\perp$ by observing the fact that the color
of the lower stripe gets stronger as $\lambda_\perp$ increases. This indicates a weaker coupling between the two MBSs, and hence the energy $E_0$ of the ABS embed inside the impurity decreases as shown in Fig. 3(d) of the main text.\\

\begin{figure}
\centering
\includegraphics[width=9cm, clip=]{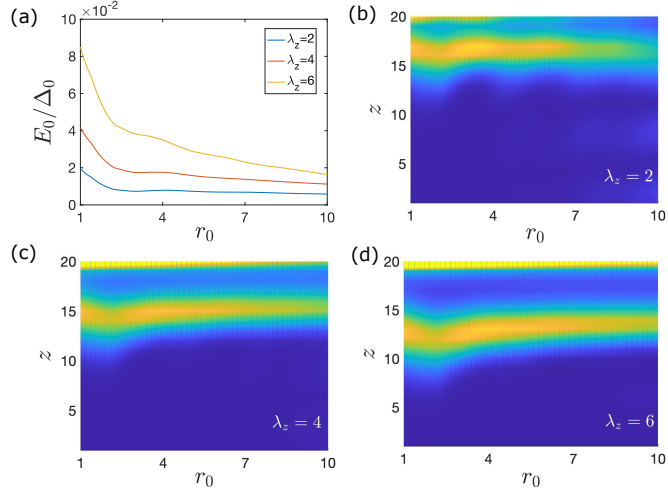}
\caption{(Color online) 
(a) The lowest energy $E_0$ as a function of the vortex radius $r_0$. (b, c, d) The LDOS $\rho_c$ along the center vortex line corresponding to the three curves in (a). Here the width of the impurity potential is fixed to $\lambda_\perp=10$. Other parameters are the same as Fig. S\ref{fig: S7}.}
\label{fig: S8}
\end{figure}

Similar to the width $\lambda_\perp$ of the impurity, the size of the SC vortex $r_0$ also has an effect on the low energy bound states. To see this, we plot the lowest energy $E_0$ as functions of $r_0$ with different $\lambda_z$ in Fig. S\ref{fig: S8}(a). The corresponding LDOSs of the MBSs are shown in Fig. S\ref{fig: S8}(b, c, d). It seems that the bottom MBS is ``squeezed" as the vortex is shrunk by observing the enhanced LDOS of the MBS. Thus, a global increase in the energy $E_0$ is observed. More interestingly, besides the energy increase, one can also observe a weak oscillation in each curve of Fig. S\ref{fig: S8}(a). This oscillation can be attributed to the oscillation of the position for the bottom MBS along $z$-direction [this can especially be seen from Fig. S\ref{fig: S8}(b)], which reveals that the horizontal tuning of the vortex can have a control on the position of bound states along the vortex line.

